\magnification=1200
\baselineskip=18truept
\def\bold1{{\bf 1}}

\def\exct{1}
\def\ovlap{2}
\def\trunc{3}
\def\geom{4}
\def\hern{5}
\def\practa{6}
\def\scria{7}
\def\practb{8}
\def\scrib{9}
\def\liu{10}
\def\su2anom{11}
\def\kikneub{12}
\def\nar{13}
\def\riken{14}

\line{\hfill RU-99-03}

\vskip 2truecm
\centerline{\bf The overlap lattice 
Dirac operator and dynamical fermions}

\vskip 1truecm
\centerline{Herbert Neuberger}
\vskip .5truecm
\centerline{\tt neuberg@physics.rutgers.edu}
\vskip 1truecm
\centerline {Department of Physics and Astronomy}
\centerline {Rutgers University, Piscataway, NJ 08855-0849}

\vskip 2truecm

\centerline{\bf Abstract}
\vskip 0.75truecm
I show how to avoid a two level nested 
conjugate gradient procedure
in the context of Hybrid Monte Carlo with
the overlap fermionic action. The resulting
procedure is quite similar to Hybrid Monte Carlo
with domain wall fermions, but is more flexible
and therefore has some potential worth exploring.

\vfill\eject

By now it is clear that strictly massless QCD
can be put on the lattice without employing
fine tuning [\exct]. At the moment, all practical
ways to do this are theoretically based
on the overlap [\ovlap,\trunc]. The 
procedures are practical only because the fermionic
matrix $D$ admits a simple expression in terms
of a function $\varepsilon$ of a very sparse matrix
$H_W$. $D$, $\varepsilon$ and $H_W$ will be defined below.

One may wonder why we need to settle for a function $\varepsilon$
of a sparse matrix, and not use a matrix $D$ which is sparse by
itself. The main point is that $\varepsilon$ is not analytic
in its argument while $H_W$ is analytic in the link gauge fields,
given by unitary matrices $U_\mu (x)$, where $\mu =1,2,3,4$ denotes
a direction and $x$ a lattice site. However, for strictly massless
fermions, $D$ cannot be analytic in the link variables. Indeed,
if it were, we could not have eigenvalues of $D$ depending nonanalytically
on the link variables: On the lattice a set of link variables
all set to unity can be smoothly deformed to a good approximation
to an instanton. In the process of this deformation
the eigenvalue of $D$ closest to zero will move until some 
intermediate set of links variables is reached 
and after that will be stuck at zero. More complicated 
evolutions can also occur, but they all have to be nonanalytic in
the link variables.
Moreover, the lack of analyticity comes from the {\it global}
structure of the gauge background described by the link variables.
A sparse and local $D$ cannot provide such an eigenvalue movement. 
In the overlap, the burden of introducing the nonanalyticity is carried by
the function $\varepsilon$. The dynamics can then be relegated to 
a sparse matrix $H_W$ which is local and analytic in the link
variables. That only a function
of a sparse matrix enters
makes it still possible to use polynomial Krylov
space methods and avoid full storage of the fermion matrix. It is
well known that full storage is prohibitive at reasonable
system sizes.  

The explicit form of the massless fermionic
matrix $D$ is
$$
D={1\over 2} (1+\epsilon^\prime\epsilon),
\eqno{(1)}
$$
where $\epsilon^\prime$ and $\epsilon$ are
hermitian and square to unity. Thus, $V\equiv
\epsilon^\prime\epsilon$ is unitary. Replacing
$1$ by a parameter $\rho$ slightly larger (smaller) than
unity corresponds to giving the fermions a
positive (negative) mass.\footnote{*}
{The pure gauge field action is assumed to have
zero theta-parameter.} Switching the sign of the physical
mass corresponds to replacing $\rho$ 
by ${1\over \rho}$ [\trunc].
$D$ acts on a space of dimension
$4Vn_c$ where $V$ is the number of lattice sites ($x$),
$4$ is the number of spinorial indices ($\alpha,\beta$) and 
the fermions are in the fundamental representation
of $SU(n_c)$, also carrying a group index $i$. 
Usually, one suppresses the spinorial and group indices,
but indicates the site indices explicitly. One uses
the indices $\mu,\nu$ both for directions and 
for vectors of length one in the respective direction. 

Clearly, $\det \epsilon = 
(-1)^{{1\over 2} tr \epsilon}$ and the same
is true of $\epsilon^\prime$. The latter is
picked so that $tr\epsilon^\prime =0$ for all
gauge fields. The simplest choice 
is $\epsilon^\prime =\gamma_5$, in which case
all the gauge field dependence comes in through
$\epsilon$. ${1\over 2} tr \epsilon$ is 
the topological charge of the background [\exct,\ovlap]. 
$\epsilon$ is defined by
$$\eqalign{&
\epsilon ={H_W\over
{\sqrt {H_W^2}}},~~H_W=\gamma_5 D_W 
= H_W^\dagger ,\cr
&(D_W\psi ) (x)=\psi(x)/\kappa -\sum_\mu [(1-\gamma_\mu )U_\mu (x)
\psi (x+\mu ) + (1+\gamma_\mu ) U^\dagger_\mu (x-\mu) \psi(x-\mu) ]
\cr} \eqno{(2)}$$
where $D_W$ is the Wilson lattice Dirac operator
with hopping $\kappa$ in the range $(.125,.25)$. The matrices
$\gamma_\mu$ are Euclidean four by four Dirac matrices acting
on spinor indices.

$\epsilon$ is defined for all gauge orbits with
$H_W^2 >0$. It is nonanalytic when $H_W$ has
a zero eigenvalue. This exclusion of the ``zero measure''
set of gauge fields where $H_W$ has exact zero modes is
necessary [\exct,\ovlap], as explained above,
in order to cut the space of lattice gauge orbits
up into different topological sectors. The space
of allowed gauge backgrounds has also to
provide a base manifold capable of supporting the
nontrivial $U(1)$ bundles needed to reproduce
chiral anomalies [\geom]. 

Saying that the space of forbidden gauge orbits has
zero measure is not really sufficient to discount
other possible consequences of the nonanalyticity of
$D$: for example, one may be worried that nonlocal
effects are somehow introduced, and the regularized
theory isn't going to become massless $QCD$ in the
continuum limit. That there are no bad side
consequences of the nonanalyticity is obvious
from the following observation:
Gauge fields with relatively small 
local curvature (in other words, with all parallel
transporters round elementary plaquettes close
to unitary matrices - note that this is a gauge
invariant requirement) will produce an $H_W^2$ bounded 
away from zero. Indeed,
the spectrum of $H_W$ is gauge invariant and has
a gap around zero on the trivial orbit. Thus,
the above is evident by continuity. (More formal
arguments have recently appeared in [\hern].) The
continuum limit is dominated by gauge configurations 
which are far from the excluded backgrounds where
$H_W$ is non-invertible. That this had to be quite
obvious follows from the fact that $D_W$, by itself,
would describe {\it massive} fermions on the lattice,
and that this mass, controlled by the variable $\kappa$,
is kept of order inverse lattice spacing $a$ when $a$
is taken to zero.

In theory we need $\epsilon =\varepsilon (H_W )$,
where $\varepsilon (x) $ is the sign function giving the
sign of $x$, and is nonanalytic at $x=0$. However,
since $H_W$ will never have exactly zero eigenvalues,
a numerical implementation of $\varepsilon (x)$
seems possible. Of course, what determines the level of difficulty
is how close the numerical implementation of $\varepsilon (x) $
has to be to the true $\varepsilon (x) $ for all
values of $x$ that are possible. The set of values of
$x$ we need to consider is the set of values the eigenvalues
of $H_W$ can take. In practice, $H_W$ can have eigenvalues
close to zero, and, since in that vicinity the true 
$\varepsilon (x)$ has a jump, the numerical implementation
inevitably becomes expensive (in cycles) for gauge fields
that produce an $H_W$ with numerically tiny eigenvalues. 
Of course, the overall scale of $H_W$ is irrelevant for 
the sign function, so it is the ratio between the largest
and smallest eigenvalues (in absolute value) that matters.
This is the condition number of $H_W$, and it plays a significant
role in all numerical considerations that follow.

The method that seems most promising is to compute
the action of $\epsilon$ on a vector $\phi$ 
by approximating the sign function
$\varepsilon (x)$ by a ratio of polynomials
$$
\varepsilon (x)\approx \varepsilon_n (x)
\equiv {{P(x)}\over {Q(x)}},\eqno{(3)}$$
where the $deg(Q)=deg(P)+1=2n$. A simple
choice, which obeys in addition 
$|\varepsilon_n (x)|<1$, is [\practa]:
$$
\varepsilon_n (x)={{(1+x)^{2n} - (1-x)^{2n}}\over
{(1+x)^{2n} + (1-x)^{2n}}}=
{x\over n} \sum_{s=1}^n {1\over
{\cos^2 [(s-{1\over 2}) {\pi\over {2n}}]x^2
+\sin^2 [(s-{1\over 2}){\pi\over {2n}}]}}.
\eqno{(4)}$$
This choice also respects the symmetry $\varepsilon (x)
=\varepsilon ({1\over x})$. Thus, it treats the
extremities of the spectrum of $H_W$ equally.
There is a certain advantage in knowing that the
approximation $\varepsilon_n (x)$ never exceeds unity
in absolute magnitude. This ensure that the related approximate
matrix $D$ never has strictly zero eigenvalues, a source of concern
when $D$ itself is inverted, something we need to also do. 

It was pointed out in [\scria] that
abandoning $|\varepsilon_n (x)|<1$,
the quantity $\max_{x\in (a,b)} |\varepsilon (x)
-\varepsilon_n (x)|$ ($a<H_W <b$) can be
minimized for fixed $n$ and the needed $n$
for a given accuracy can be reduced relative to
(4). It is advantageous to work with a small $n$.
It is not known at the moment whether the
tradeoff between a smaller $n$ and $|\varepsilon_n (x)|<1$
is beneficial. In this context let me observe that
one can always use (4) with $n=1$ on any other
sign-function approximation of $H_W$. This doubles
the effective $n$, but reintroduces $|\varepsilon_n (x)|<1$.
It does not reintroduce the inversion symmetry under
$x\to 1/x$, but
the latter may be less important in practice. 

Whichever polynomials one uses, the main point 
is that a fractional decomposition as in (4)
makes it possible to evaluate the action of
$\epsilon$ at the rough cost of a 
single conjugate gradient inversion of 
$H_W^2$ [\practa]. The parameter $n$ only affects
storage requirements, and even this 
can be avoided at the expense of an increase by
a factor of order unity in computation [\practb]. Thus,
keeping $n$ as small as possible is not necessarily
a requirement. On the other hand, for the approximation
to the sign function to be valid down to small arguments,
one shall need to pick relatively large $n$'s and face the slow-down
stemming from the single conjugate gradient inversions 
being controlled essentially by the condition number
of $H_W^2$ itself, with no help from the $n$-dependent
shift. 

In both types of rational approximants [\practa,\scria] 
there exists a polynomial $q$ of rank $n$ such that
$$
Q(x ) = |q(x)|^2 .\eqno{(5)}$$
This equation simply reflects the positivity of $Q$ 
on the real line; it makes it possible to work with
Hermitian matrices below.

In quenched simulations one needs quantities 
of the form [\trunc]:
$$
{{1-V}\over{1+V}} \phi \equiv ({1\over D} -1 )\phi=
{1\over{\epsilon^\prime + \epsilon}} ( \epsilon^\prime - \epsilon) \phi=
(\epsilon -\epsilon^\prime){1\over{\epsilon^\prime + \epsilon}}\phi.
\eqno{(6)}$$
The inversion of $D$, or $\epsilon^\prime + \epsilon$, 
needs yet another 
conjugate gradient iteration, and one
ends up with a two level nested conjugate 
gradient algorithm.  The operator that needs
to be inverted $\epsilon^\prime + \epsilon$, 
is hermitian and this is a potentially useful
property numerically. This operator, as
seen in the above equation, anticommutes with
$\epsilon^\prime - \epsilon$, and since the latter
is generically non-degenerate, has a spectrum which
is symmetric about zero.

The operators $H_\pm \equiv {1\over 2}
[\epsilon^\prime \pm \epsilon]$ have some nice properties 
so some comments about how they relate to $D$
are in order [\scria,\scrib,\su2anom]:
Since $VH_\pm V= V^\dagger H_\pm V^\dagger = H_\pm$,
$[H_\pm,V+V^\dagger]=\{H_\pm , V-V^\dagger\}=0$. Note
that $V+V^\dagger=\{\epsilon,\epsilon^\prime\}$ and
$V-V^\dagger=[\epsilon^\prime,\epsilon]=\pm 4H_\mp H_\pm$.
We also know that $K\equiv Ker([\epsilon,\epsilon^\prime])=
Ker(H_+)\oplus Ker (H_-)
=Ker(1+V)\oplus Ker (1-V)$. On the complement of $K$, $K_\perp$, 
eigenvalues of $H_+$, $h$, satisfy $0<|h|<1$  and come in
pairs $h^\pm =\pm|h|\equiv\pm\cos {\alpha\over 2}$ with
$0<\alpha<\pi$. The eigenvalues of $H_-$ are 
$\pm \sin {\alpha\over 2}$ on $K_\perp$ in the same
subspaces. Corresponding to each eigenvectors/eigenvalues
pair of $H_\pm$ are a pair of eigenvectors/eigenvalues of
$V$ (and $V^\dagger$) with eigenvalues $e^{\pm i\alpha}$. 
The two pairs of eigenvectors are linearly related. 
States relevant to the continuum limit have $\alpha\approx \pi$.
These features generalize to the massive case, where one has to deal
with $H_{ab}$ and $H_{ba}$, where the matrix pencils
are defined as $H_{ab}=a\epsilon +b\epsilon^\prime$ with real $a,b$.

To see directly why the $H_\pm$ are special we follow [\kikneub] 
and represent them using two distinct bases,
one associated with $\epsilon$ and the other
associated with $\epsilon^\prime$: $\epsilon \psi_i =\epsilon_i \psi_i$,
$\epsilon^\prime \psi^\prime_i = \epsilon^\prime_i \psi^\prime_i$.
Then $<\psi^\prime_i, H_\pm \psi_j > = 
{{\epsilon^\prime_i \pm \epsilon_j}\over 2} <\psi^\prime_i , \psi_j >$,
showing that $\det H_\pm$ factorizes [\nar] since matrix elements
corresponding to $\epsilon^\prime_i\pm\epsilon_j =0$ vanish.
Exactly half of the $\epsilon^\prime_i$ are $1$ and the rest are $-1$.
When $\epsilon$ is approximated, $|\epsilon_i|$ will no longer be 
precisely unity and we get some right-left mixing. 

Getting back to our main topic, we have ended up with a nested
conjugate gradient procedure. 
This is not prohibitive in
the quenched case [\scrib], but makes the entire
approach only tenuously feasible with present
computational resources when 
dynamical simulations using Hybrid Monte Carlo
are contemplated [\liu].

My objective here is to show that
in the context of Hybrid Monte Carlo, a
nested conjugate gradient procedure can possibly be
avoided. Of course, this comes at some cost
and only future work can tell how well the
idea works. At this stage I only wish to draw attention
to an alternative to using
a nested conjugate gradient procedure
in simulations with dynamical fermions.

As usual with Hybrid Monte Carlo, we 
work with an even number of flavors.
Obviously,
$$
\det D =\det (\epsilon^\prime D) 
=\det {{\epsilon^\prime +\epsilon}\over 2}
\approx \det {1\over 2} [\gamma_5 +\varepsilon_n (H_W ) ].\eqno{(7)}$$
But,
$$
\det {1\over 2} [\gamma_5 + \varepsilon_n (H_W )]=
{{\det {1\over 2} [q(H_W )\gamma_5 q^\dagger 
(H_W) +P(H_W ) ]}\over
{\det[Q]}}.\eqno{(8)}$$
For example, with the polynomials of (4) we
have $q(x)=(1+x)^n +i(1-x)^n$. 

The denominator $\det [Q]$ in (8) can 
be implemented by pseudofermions - by this term
I mean variables in the functional integrals that
carry the same set of indices as fermions do,
only they are bosonic, so integration over the
exponent of a quadratic form in pseudofermions is restricted
to positive kernels, and produces the inverse
of the kernel's determinant. Note that
$Q > 0$ and one does not need an even power here to
ensure positivity. 

One does need an even power nevertheless, 
because it is a requirement embedded in the Hybrid Monte
Carlo algorithm: In that algorithm, one needs to 
invert the fermion matrix, $M$, in the course of computing
the Hybrid Monte Carlo force. 
$M=q(H_W )\gamma_5 q^\dagger (H_W) +P(H_W )$ is
not positive definite, but should be - and this is achieved
by doubling the number of fermions. 
In equation (8) I chose to factor the expression in such a way
that $M$ come out hermitian. I did this because numerical
procedures are easier understood theoretically when the matrices
are hermitian, and also, because this may help to
reduce the condition number. But, there is no guarantee that
it is really beneficial to make $M$ hermitian. Therefore
let me mention that 
other factorizations are 
possible: For example, using the approximation in (4),
$$
\det {1\over 2} [\gamma_5 +\varepsilon_n (H_W )] =
{{\det \left [ {{1+\gamma_5}\over 2 } (1+H_W )^{2n}
-{{1-\gamma_5}\over 2 } (1-H_W )^{2n}\right ]}\over
{\det \left [ (1+H_W )^{2n} + (1-H_W )^{2n} \right ] }}.
\eqno{(9)}$$
The matrix in the denominator is still positive definite, but
$M$ is not hermitian now, and resembles expressions obtained
in the context of another truncation of
the overlap [\trunc], known as domain wall fermions [\riken]

The appearance of pseudofermions renders this
case even closer to so called domain wall fermions.
The trade-off is between an extra dimension
there and the higher degree polynomials here. 
In the present approach there is more 
flexibility and one does not keep unneeded
degrees of freedom in memory; still, it would
be premature to decide which approach is best.
Optimizing [\scria] 
to make $n$ as low as possible seems now again 
worthwhile, more so that in the quenched case. 

The cost of an $M\cdot \phi$ operation is roughly $4n$
times the cost of an $H_W\cdot \phi$ operation. The
condition number of $M$ may also be 
larger than that of $H_W$ and 
increase with $n$. It is therefore important to find out
what the smallest $n$ one can live with is. It could be that
it turned out to be too hard to maintain $\varepsilon_n (x)$
a good approximation to $\varepsilon(x)$ while keeping the
condition number of $M$ manageable. If one focuses only on
the quality of the approximation to the sign function it 
is actually likely that the condition number of $M$
will be large\footnote{*}{R. Edwards, private communication.}
because of the high degrees of the polynomials:
Consider $\psi$, a normalized eigenstate of $H_W$ with eigenvalue
$h$. We find $\psi^\dagger M \psi = P(h) +Q(h) \psi^\dagger \gamma_5 \psi$.
Both $P(h)$ and $Q(h)$ can be very big numbers (for large degree
$n$). In absolute magnitude they are very close, this is why
the ratio $P(h)/Q(h)$ is close to $\pm 1$. But, this 
cancelation can be easily spoiled by the 
$\psi^\dagger \gamma_5 \psi$ factor, and thus $M$ can have very large
eigenvalues. There is little reason to hope for $M$ to have no
small eigenvalues, so it might be the case that $M$ has unacceptable
large condition numbers when $n$ is too large.

I now wish to show that one can try to avoid 
this latter problem by introducing extra fields. This is,
I believe, the essential reason why 
domain wall fermions [\trunc,\riken] are at all practical. 

To understand this additional trick we start from some
relatively easily proven identities.
Consider a fermionic bilinear action $S_0$:
$$
S_0 =\bar\psi\gamma_5\psi +\bar\psi\bar A_1\phi_1 -\bar\phi_1 A_1 \psi +
\phi_1 B_1 \phi_1 + \dots + \bar\phi_{n-1} \bar A_n \phi_{n} -
\bar\phi_n A_n \phi_{n-1} +
\bar\phi_n B_n \phi_n .\eqno{(10)}$$
The fields with bars are rows, the ones without are columns and $A_i ,
\bar A_i , B_i$ are commuting matrices. One can visualize this
action as a chain, extending into a new dimension. The degrees
of freedom we are interested in sit at one end of the chain; these
are the $\bar\psi ,\psi$ fields. The $\bar\phi ,\phi$ fields are
the extra fields I introduced to handle the condition number problem.
The idea is to arrange matters so that integrating out all the
$\bar\phi , \phi$ variables will produce an action for the variables
$\bar\psi ,\psi$ of the precise rational form we wish. But, the
condition number that will be relevant numerically, 
will be the condition number 
of the bigger kernel in $S_0$, involving all fermionic fields. 

To get the induced action for the fields $\bar\psi,\psi$ we integrate
out the fields $\bar\phi, \phi$ starting from the other end
of the chain. The integration over the pair of fermions at the
end of the chain produces a factor of $(\det B_n )$ in front
and adds a piece to the quadratic term $B_{n-1}$, 
coupling $\bar\phi_{n-1}$ to $\phi_{n-1}$, of the form $A_n \bar A_n / B_n $.
Now this can be iterated until the last pair of $\bar \phi ,\phi$
is integrated out. We have obtained the following identity:

$$
\int d\bar\phi_1 d\phi_1 \dots d\bar\phi_n d\phi_n e^{S_0} =
\prod_{i=1}^n (\det B_i ) e^{\bar\psi (\gamma_5 +R )\psi },
\eqno{(11)}$$
where,
$$
R={ {\displaystyle A_1\bar A_1}\over \displaystyle B_1 +
                 {\strut {\displaystyle A_2\bar A_2 } \over\displaystyle B_2 +
                   {\strut {\displaystyle A_3\bar A_3} \over\displaystyle B_3 +
                 \dots    {\strut \displaystyle  \ddots \over \displaystyle B_{n-1} +
                     {\strut {\displaystyle A_n \bar A_n } \over
    \displaystyle B_n }}}}}
\eqno{(12)}$$

Now, the expression for $R$ is recognized as a truncated
continued fraction. Any ratio of polynomials can
be written as a truncated continued fraction by 
the Euclid algorithm (invert, divide with remainder in
the denominator and continue). Thus, we learn that any
fractional approximation we wish to use for the sign
function can be mapped into a chain with only nearest neighbor
interactions.

Usually, $P$ in (3) is odd and $Q$ is even:
$P(x)=xP_1(x^2),~Q(x)=Q_1(x^2)$. $P_1$ is of degree $n-1$ and
$Q_1$ is of degree $n$. Therefore, the truncated continued fraction
has the following structure:
$$
{{P_1 (u)}\over {Q_1 (u)}}=
{ {\displaystyle \alpha_0 }\over \displaystyle u +\beta_1  +
                 {\strut {\displaystyle \alpha_1 } \over\displaystyle 
                                      u+\beta_2 +
                   {\strut {\displaystyle \alpha_2 } \over\displaystyle 
                                      u+\beta_3 + \dots
                   {\strut \displaystyle   \ddots \over 
{\displaystyle u+\beta_{n-1}}+
                     {\strut {\displaystyle \alpha_{n-1}} \over 
{\displaystyle u+\beta_n} }}}}}\eqno{(13)}$$
Picking $B_i = H_W^2 +\beta_i,~i=1,2,\dots,n$ 
and $A_1\bar A_1 =\alpha_0 H_W , A_i\bar A_i =\alpha_{i-1} ,~
i=2,3\dots,n$ produces the desired expression. Again, one needs
pseudofermions to compensate for the prefactor $\prod \det (H_W^2 +\beta_i )$
in (11). For the kernels of the
pseudofermions to be positive definite we need $\beta_i \ge 0$.

A similar trick can be used if one wants to implement ${{P_1 (u)}\over
{Q_1 (u)}} = \sum_{i=1}^n {{\alpha_i^2}\over {u+\beta_i^2}}$.
Now $S_0 = \bar\psi \gamma_5 \psi +\sum_i\alpha_i (\bar\psi H_W 
\phi_i + \bar\phi_i \psi ) -\sum_i \bar\phi_i (H_W^2
+\beta_i^2 )\phi_i $. Again, one needs pseudofermions. The structure
is somewhat different.

One can avoid having the squares of $H_W$ in the chain by more
continued fraction expansion. As an example, I worked out
the explicit map of the approximation
$\varepsilon_n (x)$ of equation (4) to a chain involving only $H_W$
terms (which are the least expensive to implement). To start,
I use a formula that goes as far back as Euler: 
$$
\varepsilon_n (x) =
{ {\displaystyle 2nx }\over \displaystyle 1 +
                 {\strut {\displaystyle (4n^2-1)x^2 } 
\over\displaystyle 
                                      3+
                   {\strut {\displaystyle (4n^2-4)x^2 } 
\over\displaystyle 
                                      5 + \dots
                   {\strut \displaystyle   \ddots \over 
{\displaystyle 4n-3}+
                     {\strut {\displaystyle 
[4n^2 - (2n-1)^2]x^2} \over 
{\displaystyle 4n-1} }}}}}\eqno{(14)}$$

Now, I use invariance under inversion of $x$ to move the $x$ factors
around. I also change some signs to make the expression more 
symmetrical. The end result for the path integral is:
$$
\int d\bar\phi_1 d\phi_1 \dots d\bar\phi_n d\phi_n e^{S_*} =
(\det H_W )^{2n} e^{\bar\psi (\gamma_5 +\varepsilon_n (H_W ) )\psi },
\eqno{(15)}$$
To write down the quadratic action $S_*$ we introduce the
extended fermionic fields $\bar \chi , \chi$:
$$
\bar \chi = \pmatrix {\bar\psi & \bar\phi_1 & \dots & \bar\phi_{2n}\cr},~~~
\chi=\pmatrix {\psi & \cr\phi_1 & \cr\vdots &\cr \phi_{2n}}\eqno{(16)}$$
$S_* = \bar\chi {\bf H} \chi$ where the new kernel, ${\bf H}$, in block
form, has the following structure:
$$
{\bf H} =\pmatrix {\gamma_5 & \sqrt {\alpha_0 }& 0& 0& \dots &\dots & 0 \cr
          \sqrt {\alpha_0 }& H_W & \sqrt {\alpha_1 }& 0&\dots &\dots &0\cr
           0& \sqrt {\alpha_1 } & -H_W & \sqrt {\alpha_2 }&\dots&\dots & 0\cr
                  \dots & \dots & \dots & \dots &\ddots & \dots & 0\cr
                   \dots & \dots & \dots & \dots &\dots & H_W & \sqrt {
                                           \alpha_{2n-1}}&\cr
	          \dots & \dots & \dots & \dots &\dots & \sqrt {
                   \alpha_{2n-1}}& -H_W \cr}\eqno{(17)}$$
The numerical coefficients $\alpha$ are given below:
$$
\alpha_0 =2n ,~ ~~\alpha_j = {{(2n-j)(2n+j)}\over {(2j-1)(2j+1)}},~ j=1,2,...
\eqno{(18)}$$

The main point is that the structure of the extended kernels is sufficiently
close to five dimensional fermions that we can be quite sure that the condition
numbers are similar to the ones encountered with domain wall fermions.

Note that we now have a hermitian kernel, ${\bf H}$. This would be useful
if we wanted to use Lanczos techniques to study the entire
eigenvalue spectrum of ${\bf H}$. Actually, to do this in an
efficient way (i.e. applying the Cullum-Willoughby
method) one needs full 64 bit precision, to be able to distinguish so called
spurious eigenvalues from true ones. One cannot get the needed
accuracy if one uses a direct evaluation of the action of $\gamma_5 D$. 

One may wonder whether the law of conservation of difficulties
is not violated: how can it be that we compute the {\it same}
$\bar\psi \psi$ propagator ($ {1\over{\gamma_5 + R}}$ is given
by the $\bar\psi$-$\psi$ block in ${\bf H}^{-1}$ )
both in the pole method and in the
extra fields method, and use conjugate gradient algorithms (essentially)
in both cases, but hope for big gains  in one method relative to the
other ? At its essence the new trick is algorithmical - basically,
the space in which the conjugate gradient operates is enlarged
when extra fermions are added, and thus, some of the difficulties
one encounters in the smaller space are avoided. Very roughly,
what the extra fermions do, is to provide ways around
barriers one faces in the original space. This is
somewhat similar to solving the problem of minimizing a function over
a discrete set, by first making the set continuous.

Clearly,
$n$ plays a role analogous to the size of the extra dimension,
$N$, in domain wall simulations. Thus, these two truncations
of the overlap may end up being similar not only
conceptually, but also numerically. The derivation of 
our identities went through essentially
the same steps as those employed in [\trunc] only now in reverse order.
When comparing to domain wall fermions it becomes apparent that
now we can work with a hermitian kernel, that we have
much more flexibility, and that it is probably possible to
exploit any efficient, possibly 
computer architecture dependent, implementation of the action
of the lattice Wilson-Dirac operator $D_W$. 
I have not separated the chiral components of $\bar\psi,\psi$ here,
something that is quite natural in the domain wall viewpoint, [\trunc].
Of course, if there is an efficiency reason, one could try to separate
the chiral components in the more general framework presented here, too.

As always, if any single trick is useful, one usually considers
possible combinations. By this I mean to exploit both the
direct approach based on a sum of pole terms and the indirect
approach based on extra fields. The essence of the numerical
problem is that we wish to contract the spectrum of $H_W$ 
to two points, $\pm 1$. The negative half of the spectrum
of $H_W$ is intended to map into $-1$ and the positive
half to $+1$. Any map that reduces the ranges of the negative
and positive halves of the spectrum is useful. It produces
a new $H_W$ that can be used as an argument for a new map,
which now can work easier. In this way we try to combine
the good properties of various maps. Moreover, all that
$H_W$ needs to be, an this is important, is a reasonable
discretization of the continuum Dirac operators with a {\it negative}
mass of order inverse lattice spacing. Actually, one
can add as many fermion fields with large {\it positive }
masses as one wishes. Thus, if we used a few extra fields
to produce an effective $H_W$ (say for the end of the chain)
which had some reasonable spectral properties, we could,
conceivably, by
adding a parameter $\rho$ (as discussed at the beginning of this
paper) make the mass of one fermion negative and use this
action as the argument of a rational approximation
implemented by the pole method. A better behaved input would
be much easier to handle. Or, we could reverse the procedures.
For example assume you wish to use a very short chain, say
consisting only of  $\bar\psi \psi$ and $\bar\phi \phi$ with
an action (employing previous notation) $\bar\chi {\bf H} \chi$
given by:
$$
{\bf H} =\pmatrix {\gamma_5 & (H_W^2 )^{1/4} \cr
                    (H_W^2 )^{1/4} & -H_W \cr}\eqno{(19)}$$
Since the quantity $( H_W^2 )^{1/4}$ is less violently behaved around zero
than $\varepsilon (H_W )$ 
a rational approximation (or maybe even just a polynomial approximation) 
might be quite manageable. 

My main message in this paper is that in the context of dynamical
fermion simulations there are many alternatives and
tricks that have not been yet explored, and it might be
a waste to exclusively focus on the most literal numerical 
implementations of the recent 
theoretical advances on the topic of chiral symmetry on the lattice.

{\bf Acknowledgments:} This research was 
supported in part by the DOE under grant \#
DE-FG05-96ER40559. Thanks are due to R. Edwards, C.
Rebbi and to P. Vranas for comments that helped and
motivated me to write this note. 

\vskip .1truecm

{\bf  References:}

\item{[\exct]} H. Neuberger, Phys. Lett. B417 (1997) 141;
Phys. Lett. B427 (1998) 125.
\item{[\ovlap]} R. Narayanan, H. Neuberger, Phys. Lett. B302 (1993) 62; 
Nucl. Phys. B412 (1994) 574; Phys. Rev. Lett. 71 (1993) 3251;
Nucl. Phys. B443 (1995) 305.
\item{[\trunc]} H. Neuberger, Phys. Rev. D57 (1998) 5417. 
\item{[\geom]} H. Neuberger, hep-lat/9802033. 
\item{[\hern]} P. Hernandez, K. Jansen, M. L{\" u}scher,
hep-lat/9808010, C. Adams, hep-lat/9812003.
\item{[\practa]} H. Neuberger, Phys. Rev. Lett. 81 (1998) 4060. 
\item{[\scria]} R. Edwards, U. Heller, R. Narayanan,
hep-lat/9807017. 
\item{[\practb]} H. Neuberger, hep-lat/9811019.
\item{[\scrib]} R. Edwards, U. Heller, R. Narayanan,
hep-lat/9811030. 
\item{[\liu]} Chuan Liu, hep-lat/9811008.
\item{[\su2anom]} H. Neuberger, Phys. Lett. B437 (1998) 117.
\item{[\kikneub]} Y. Kikukawa, H. Neuberger, A. Yamada, Nucl. 
Phys. B526 (1998) 572.
\item{[\nar]} R. Narayanan, Phys. Rev. D58 (1998) 097501.
\item{[\riken]} 
D. B. Kaplan, Phys. Lett B288 (1992) 342;
D. Boyanowski, E. Dagotto, E. Fradkin, Nucl. Phys. B285 (1987) 340;
Y. Shamir, Nucl. Phys. B406 (1993) 90;
P. Vranas, Phys. Rev. D57 (1998) 1415;
P. Chen et. al. hep-lat/9809159;
T. Blum, A. Soni,  Phys. Rev. D56 (1997) 174;
J.-F. Lagae, D. K. Sinclair, hep-lat/9809134.

\vfill\eject\end